\documentclass[runningheads]{llncs}

\usepackage{eccv}

\usepackage{eccvabbrv}

\usepackage{graphicx}
\usepackage{booktabs}

\usepackage[accsupp]{axessibility}  

\usepackage{hyperref}

\usepackage{orcidlink}

\begin{document}


\title{CoSyncDiT: Cognitive Synchronous Diffusion Transformer for Movie Dubbing}

\titlerunning{Abbreviated paper title}

\author{Gaoxiang Cong\inst{1,2} \and
Liang Li\inst{1} \and
Jiaxin Ye\inst{3} \and
Zhedong Zhang\inst{4} \and
Hongming Shan\inst{3} \and
Yuankai Qi\inst{5} \and
Qingming Huang\inst{2}}

\authorrunning{G.Cong et al.}

\institute{Institute of Computing Technology, Chinese Academy of Sciences, Beijing, China\\
\and
University of Chinese Academy of Sciences, Beijing, China\\
\and 
Fudan University, Shanghai, China\\
\and 
Hangzhou Dianzi University, Hangzhou, China\\
\and 
Macquarie University, Sydney, Australia\\}

\maketitle

\begin{abstract}
    Movie dubbing aims to synthesize speech that preserves the vocal identity of a reference audio while synchronizing with the lip movements in a target video. 
    Existing methods fail to achieve precise lip-sync and lack naturalness due to explicit alignment at the duration level. 
    While implicit alignment solutions have emerged, they remain susceptible to interference from the reference audio, triggering timbre and pronunciation degradation in in-the-wild scenarios. 
    In this paper, we propose a novel flow matching-based movie dubbing framework driven by the \textbf{Co}gnitive \textbf{Sync}hronous \textbf{Di}ffusion \textbf{T}ransformer (\textbf{CoSync-DiT}), inspired by the cognitive process of professional actors. 
    This architecture progressively guides the noise-to-speech generative trajectory by executing acoustic style adapting, fine-grained visual calibrating, and time-aware context aligning. 
    Furthermore, we design the Joint Semantic and Alignment Regularization (JSAR) mechanism to simultaneously constrain frame-level temporal consistency on the contextual outputs and semantic consistency on the flow hidden states, ensuring robust alignment. 
    Extensive experiments on both standard benchmarks and challenging in-the-wild dubbing benchmarks demonstrate that our method achieves the state-of-the-art performance across multiple metrics. 
  \keywords{Movie dubbing \and Flow matching \and Visual voice cloning}
\end{abstract}

\section{Introduction}
\label{sec:intro}

\begin{figure}[tb]
  \centering
  \includegraphics[width=0.95\linewidth]{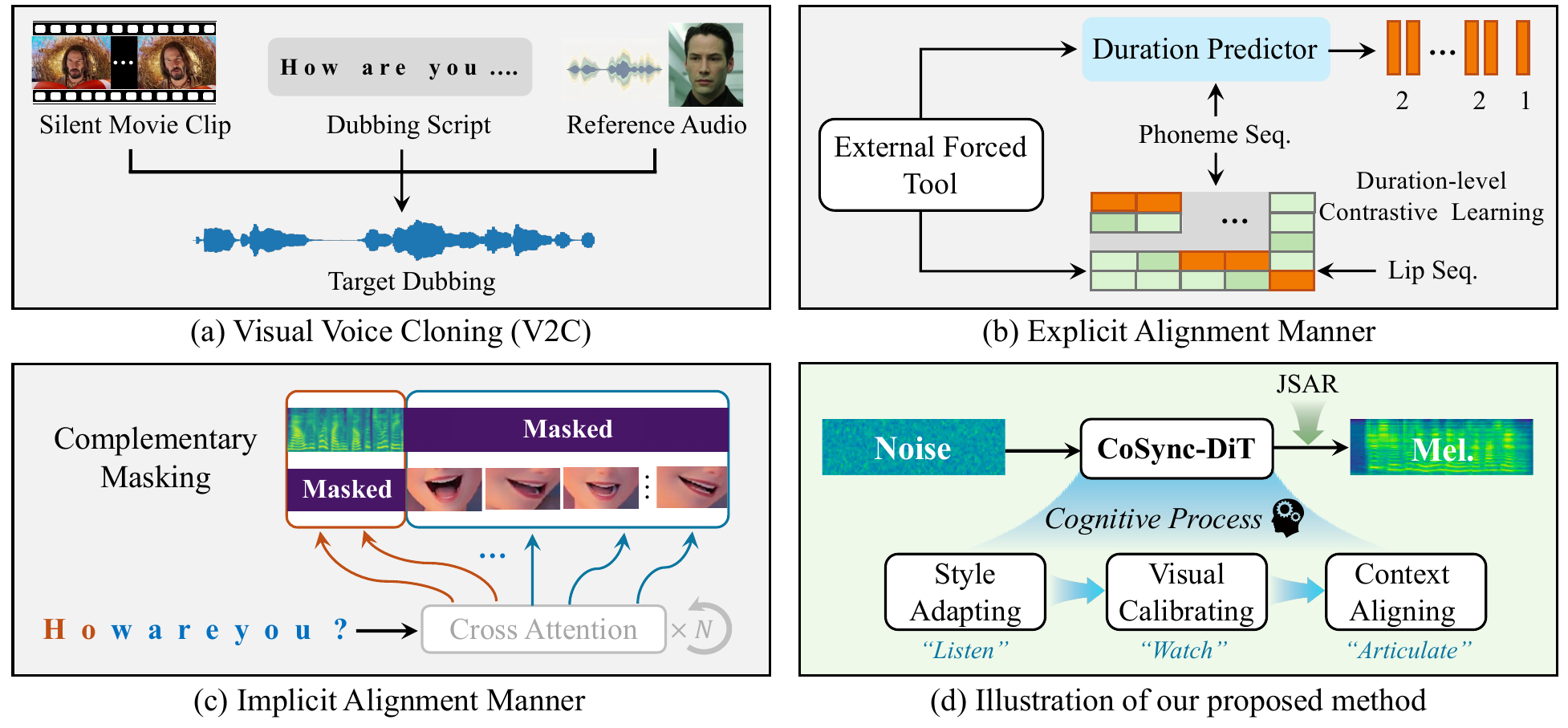}
  \caption{
  (a) Illustration of Visual Voice Cloning (V2C) task. 
  (b) Explicit alignment manner requires an external forced tool to compute the ground-truth duration of each phoneme in advance. 
  (c) Implicit alignment eliminates the dependency on external forced tools. 
  (d) We propose a movie dubbing architecture built upon CoSync-DiT, which is structured around a cognitively inspired listen–watch–articulate paradigm to progressively guide the denoising trajectory in flow matching. 
  }
  \label{fig:intro}
\end{figure}

Movie Dubbing, also known as Visual Voice Cloning (V2C)~\cite{chen2022v2c}, aims to generate a piece of speech for the character in silent videos based on the specified reference voice and textual scripts (as shown in Fig.~\ref{fig:intro}(a)). 
It promises significant potential in real-world applications such as film post-production, media production, and personal speech AIGC. 
Compared to traditional video-to-speech tasks~\cite{kim2025faces,Intelligible:conf/interspeech/ChoiKR23,HoonLet,MinsuLip}, movie dubbing presents significantly greater challenges. 
Unlike merely inferring speech from visual cues, V2C must faithfully clone the reference acoustic timbre and accurately articulate the textual scripts aligned with dynamic lip motions in the video, while ensuring emotionally expressive and high-fidelity speech quality.

Previous works mainly focus on improving pronunciation clarity and prosody modeling. 
For instance, Speaker2Dubber~\cite{zhang2024speaker} introduces a two-stage architecture by pre-training on the clean TTS corpus to improve clarity. 
Then, ProDubber~\cite{zhang2025produbber} employs a prosody-enhanced pre-training paradigm on the larger corpus. 
Recently, InstructDubber~\cite{zhang2025instructdubber}  integrates pre-trained TTS framework with LLM-driven emotion instructions to predict pitch and energy variations. 
However, these methods rely on TTS architecture and predict integer-scaled durations for each phoneme (as shown in Fig.~\ref{fig:intro}(b)), failing to achieve the precise lip-sync required for dubbing. 
Some dubbing methods~\cite{cong2024emodubber, GaoxiangFlowDubber} attempt to introduce duration-level contrastive learning between phoneme and lip sequences to improve alignment. 
Nevertheless, these methods still depend on an external forced tool (\eg, Montreal Forced Aligner~\cite{mcauliffe2017montreal}) to extract ground-truth duration boundaries for identifying positive samples during contrastive learning. 
This rigidly restricts the pronunciation to predefined intervals, damaging the naturalness and expressiveness of the generated speech~\cite{jiang2025megatts}.

Recently, AlignDiT~\cite{AlignDiT} has represented a significant leap forward in the field by eliminating the need for external forced alignment tools and adopting an implicit alignment modeling approach. 
As shown in Fig.~\ref{fig:intro}(c), it fuses text with complementary-masked audio and visual features via cross-attention across every layer of the diffusion transformer. 
However, this strategy compels the text to simultaneously align all signals, rendering the overall alignment highly susceptible to interference from the early segments of the reference audio, particularly during an in-the-wild scenario. 
Furthermore, uniformly applying text cross-attention across all DiT layers can complicate the alignment dynamics, potentially compromising both timbre and pronunciation clarity.

In this paper, we propose a novel flow matching-based movie dubbing framework built upon \textbf{Co}gnitive \textbf{Sync}hronous \textbf{Di}ffusion \textbf{T}ransformer (\textbf{CoSync-DiT}), which progressively guides a noise-to-speech generative trajectory to effectively resolve the demands of high-fidelity style cloning and exact audio-visual synchronization. 
Inspired by the dubbing workflow of professional actors~\cite{alburger2023art}, our method is structured around a sequential cognitive process of listening, watching, and articulating. 
Rather than treat all transformer layers homogeneously, CoSync-DiT partitions the denoising generation within flow matching into three distinct phases (as shown in Fig.~\ref{fig:intro}(d)). 
First, we initialize a unified acoustic-semantic prior. 
Then, the model leverages multi-head self-attention to capture the acoustic style and establish its relationship with the text, without any visual interference. 
Second, we inject fine-grained visual information through residual connections using a zero-initialized learnable gate. 
This step calibrates the previously acquired acoustic representations to capture temporal dynamics. 
Third, we design a time-aware context aligning layer to yield correct articulation by implicitly retrieving the linguistic context based on second phase. 
Finally, we introduce a Joint Semantic and Alignment Regularization (JSAR) mechanism to further rectify the third phase. 
This approach innovatively applies frame-level contrastive learning to the queried contextual outputs to enforce strict temporal alignment, while ensuring semantic constraints on the DiT hidden states.

{\bf Our key contributions are}:
{\bf (1)} We propose CoSync-DiT, a novel flow matching-based framework for movie dubbing. By formulating the denoising generation as a cognitively synchronous process, it progressively guides the vector field estimation through three sequential phases: acoustic style adapting, fine-grained visual calibrating, and time-aware context aligning. 
{\bf (2)} We design the JSAR mechanism to enforce frame-level temporal consistency on intermediate context outputs and semantic consistency on final flow hidden states, thereby stabilizing the generative trajectory and enhancing context alignment. 
{\bf (3)} Our method performs favourably against state-of-the-art methods on three public datasets, including the challenging in-the-wild movie dubbing scenarios.

\section{Related Work}

\subsection{Visual Voice cloning}
V2C~\cite{chen2022v2c, nguyen2026diflowdubber, zhao2024mcdubber, liu2025towards} has recently attracted widespread attention and promises a profound impact on film production~\cite{liu2026funcineforge, zheng2025deepdubber, tian2025dualdub, hicodit:conf/cvpr/Ye} and digital media~\cite{sung2025voicecraft, demoface:conf/icml/Ye, depmamba:conf/icassp/YeZS25, choi2024av2av}. 
V2C aims to generate synchronized speech for silent videos conditioned on a specific reference timbre. 
Early methods widely adopt explicit duration prediction to achieve lip synchronization. 
For instance, StyleDubber~\cite{cong2024styledubber} queries lip motions using textual phonemes to construct a phoneme-level visual sequence and predicts the duration of each phoneme. 
ProDubber~\cite{zhang2025produbber} and InstructDubber~\cite{zhang2025instructdubber} follow a similar paradigm. 
By enforcing explicit duration boundaries, these approaches typically guarantee high pronunciation clarity. 
However, they inherently struggle to achieve precise lip synchronization because the predicted durations must be expanded into rigid integer multiples. 
Recent methods~\cite{cong2024emodubber,GaoxiangFlowDubber} introduce duration-level contrastive learning to alleviate this limitation. 
Nevertheless, they still rely on external forced aligners to extract duration intervals. 
In this paper, drawing inspiration from the human cognitive workflow, we propose CoSync-DiT, a novel implicit dubbing framework that completely eliminates the dependency on external aligners. 
It ensures high-fidelity timbre cloning and accurate lip synchronization, even in challenging in-the-wild dubbing scenarios.

\subsection{Flow Matching and Generative Modeling} 
Flow matching~\cite{lipman2022flow} has emerged as a dominant paradigm in generative modeling due to its superior quality and inference speed. 
It learns a vector field~\cite{AlexanderCFM} to transport samples from a simple prior distribution to the target data distribution along efficient probability paths. 
In TTS field, flow matching has found extensive applications~\cite{YiweiVoiceFlow, machaTTS, SungwonFlow, VoiceboxMatthew, zhang-etal-2025, du2024cosyvoice2, yao2024stablevc, chen2024f5, zhou2025indextts2}. 
However, these methods cannot parse visual scenes and fail to achieve fine-grained lip-sync, hindering their application in movie dubbing. 
FlowDubber~\cite{GaoxiangFlowDubber} introduces a dual contrastive learning, which relies on external forced aligners and employs a monotonic search algorithm for sequential expansion. 
Nevertheless, even minor initial misalignments tend to compound throughout the generation process and  degrade the lip-sync quality. 
AlignDiT~\cite{AlignDiT} implicitly models this alignment by incorporating cross-attention modules across all transformer layers. 
However, due to its complementary masking strategy, the model struggles to simultaneously align with the reference audio and the target lip motions. 
This homogeneous injection across all layers further exacerbates alignment instability in complex dubbing scenarios. 
In this paper, we propose CoSync-DiT to progressively guide the  denoising trajectory by executing acoustic style adapting, fine-grained visual calibrating, and time-aware context aligning. 
Furthermore, we design the JSAR mechanism to learn temporal consistency and semantic consistency.

\section{Methodology} 
\begin{figure}[tb]
  \centering
  \includegraphics[width=1.0\linewidth]{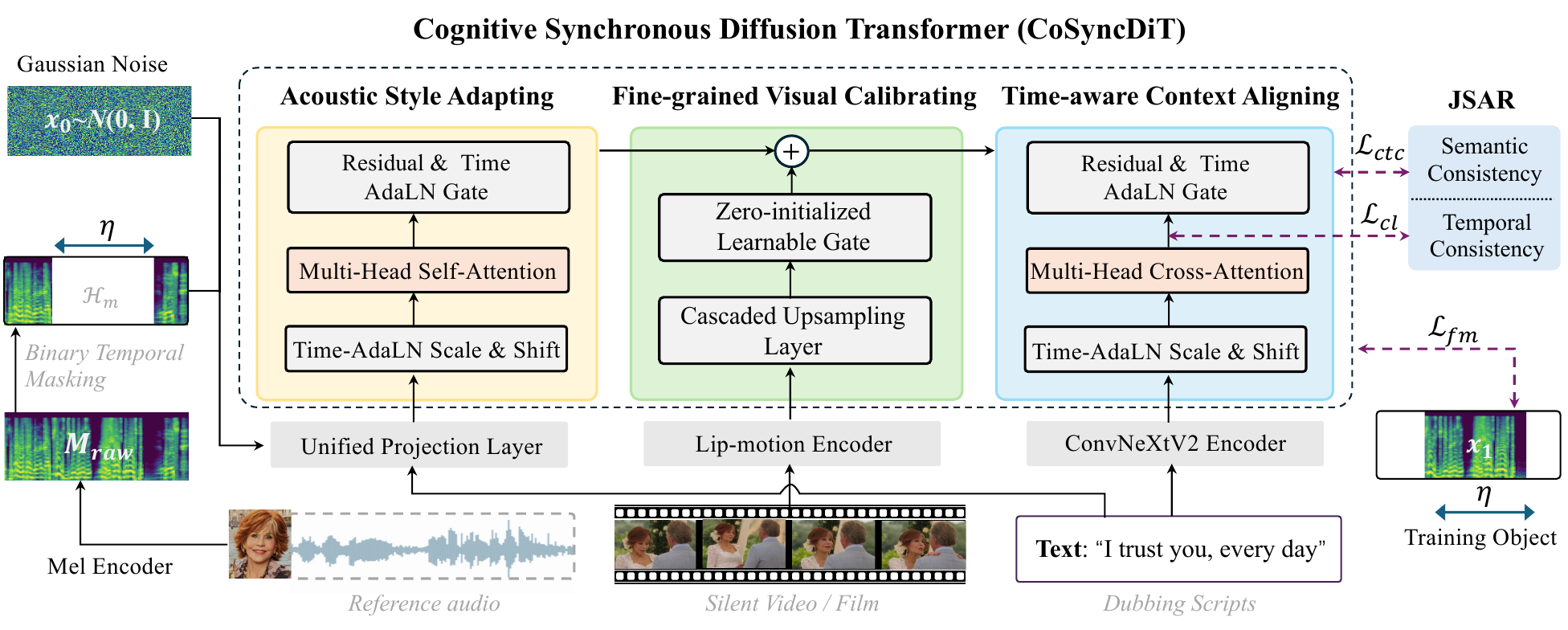}
  \caption{
  The overview of architecture 
  }
  \label{fig:archi}
\end{figure}

As shown in Fig.~\ref{fig:archi}, the overall framework of our method includes the following parts. First, individual encoders extract the foundational representations from the reference audio, silent video, and dubbing scripts. 
Next, under the Optimal-Transport Conditional Flow Matching (OT-CFM) paradigm, the proposed CoSync-DiT first takes Gaussian noise, masked acoustic conditions, and semantic conditions as priors to execute Acoustic Style Adapting. 
Sequentially, the Fine-grained Visual Calibrating focuses on capturing the rhythmic dynamics of lip motion. 
Next, a Time-aware Context Aligning module integrates the textual features to ensure precise audio-visual synchronization. 
Finally, the designed JSAR mechanism regularizes the alignment stage through semantic and temporal consistency constraints, guiding the generative trajectory toward synchronized and high-fidelity speech.

\subsection{Acoustic and Semantic Prior}
\label{sec:pre}

The reference audio $R_{a}$ is extracted to a raw mel-spectrogram $m_{raw} \in \mathbb{R}^{F \times L}$, where $F$ is the mel dimension and $L$ is the sequence length. 
A binary temporal mask $\mathbf{M}_b \in \{0, 1\}^{L}$ is then applied to obscure the target region $\eta$. 
This enables the model to yield the masked acoustic features $\mathcal{H}_{m}= (1 - \mathbf{M}_b) \odot m_{raw}$. 
Unlike directly concatenating the masked visual features, we construct a unified sequence by jointly modeling text and $\mathcal{H}_{m}$. 
To expand the word-level text sequence to the mel-level, we adopt two expansion strategies: (1) a padding operation to preserve the complete linguistic content; (2) a cross-attention operation to provide coarse temporal priors. 
Finally, we concatenate these textual sequences and $\mathcal{H}_{m}$ into the semantic-acoustic prior, rather than using raw visual signals for early-stage conditioning.

\subsection{CoSync-DiT}
\label{sec:dit}

Traditional Diffusion Transformers (DiTs) struggle to meet the strict demands of movie dubbing. 
They fail to maintain precise audio-visual synchronization, degrading both timbre fidelity and pronunciation clarity. 
Drawing inspiration from the human dubbing cognitive process, we propose the Cognitive Synchronous Diffusion Transformer (CoSync-DiT). 
The model learns to progressively establish the reference speaker’s style, perform fine-grained lip calibration, and enforce contextual alignment.

\textbf{Stage 1: Acoustic Style Adapting.} 
In the scope of OT-CFM, our model predicts the vector field mapping from a standard Gaussian noise distribution $x_0 \sim \mathcal{N}(0, I)$ to the target speech latent $x_1$, following the time-dependent path $x_t = (1 - t)x_0 + t x_1$. 
In stage 1, the model concatenates the noisy speech $x_t$ and the semantic-acoustic prior by unified projection layer to form the initial sequence $\mathcal{Z}^0$. 
The Multi-Head Self-Attention (MHSA) models the long-range dependencies in the initial hidden states to capture style information: 
\begin{equation}\mathcal{Z}^{l}_{style} = \mathcal{Z}^{l-1} + \alpha_{1}^l(t) \odot \mathrm{MHSA} ( \mathrm{AdaLN}_{\beta_1,\gamma_1}(\mathcal{Z}^{l-1}, t)),
\end{equation}
where $l$ represents the current block level. $\mathrm{AdaLN}_{\beta_1,\gamma_1}$ applies Time-Adaptive Layer Normalization (Time-AdaLN) to modulate the input feature statistics via scale $\gamma_1$ and shift $\beta_1$ vectors driven by the current time $t$. 
The gating term $\alpha_1(t)$ controls the residual contribution. 
Similarly, a time-adaptive Multilayer Perceptron (MLP) further refines these features via non-linear dimensional transformations to consolidate the relationship between  acquired acoustic style and  text.

\textbf{Stage 2: Fine-grained Visual Calibrating.} 
The input silent video is first processed by a lip-motion encoder to extract the raw lip features $\mathcal{X}_{raw}$. 
Then, $\mathcal{X}_{raw}$ is then passed through a cascaded upsampling layer to yield the refined visual representations $\mathcal{X}_{lip}$, ensuring their temporal resolution strictly matches that of the target mel-spectrogram. 
Next, $\mathcal{X}_{lip}$ are seamlessly injected into the $\mathcal{Z}^{l}_{style}$ via a zero-initialized learnable gate: 
\begin{equation}\mathcal{Z}^{l}_{lip} = \mathcal{Z}^{l}_{style} + \Lambda^l \odot \mathcal{X}_{lip}, \quad \text{where } \Lambda^l = \mathbf{0} \in \mathbb{R}^{d}.
\end{equation}
This gating mechanism $\Lambda^l$ ensures that the visual injection acts as a subtle rhythmic residual, providing fine-grained frame-level calibration while protecting previously established style information.

\textbf{Stage 3: Time-aware Context Aligning.} 
We place the Multi-Head Cross-Attention (MHCA) at the bottom of the architecture to ensure that text alignment operates on fully mature visual-acoustic representations, rather than forcing a premature and unstable complementary fusion. 
Furthermore, we introduce a time-aware mechanism, $\mathrm{AdaLN}_{\beta_3,\gamma_3}$, to modulate the cross-attention process, allowing the alignment dynamics to better adapt to the evolving flow states: 

\begin{equation}\mathcal{Z}^{l}_{out} = \mathcal{Z}^{l}_{lip} + \alpha_{3}^l(t) \odot ( \mathrm{MHCA} ( \mathrm{AdaLN}_{\beta_3,\gamma_3}(\mathcal{Z}^{l}_{lip}, t), \mathcal{H}_{text})),
\label{context_MHCA}
\end{equation}
where $\alpha_{3}^l(t)$ controls the residual contribution of the cross-attention output according to the current timestep $t$. 
The textual representation $\mathcal{H}_{text}$ extracted by ConvNeXtV2 encoder~\cite{woo2023convnext} serves as both the Key and Value, encouraging the generative path to converge toward the aligned content by implicitly retrieving the linguistic context.

\subsection{Joint Semantic and Alignment Regularization}
\label{sec:aligntwo}
While the OT-CFM efficiently constructs the data generation path, the unconstrained vector field estimation often leads to temporal misalignments. 
In this work, we introduce the Joint Semantic and Alignment Regularization (JSAR) mechanism, constraining both the final flow hidden states and the intermediate context representations.

\textbf{Alignment Regularization.} 
The intermediate context representations (\ie, the output of the MHCA in Eq.~\eqref{context_MHCA}, denoted as $\hat{\mathcal{Z}}_{ca}$) need to maintain inherent temporal consistency when queried by the flow hidden features. 
To implicitly align these context representations in time, we introduce a frame-level contrastive learning by employing the audio branch representations from a pre-trained AV-HuBERT (denoted as $\mathcal{F}_{av}$) as the contrastive ground truth. 
Both the outputs $\hat{\mathcal{Z}}_{ca}$ and the AV-HuBERT features $\mathcal{F}_{av}$ are then L2-normalized along the feature dimension. 
Then, it uses the InfoNCE objective to maximize the cosine similarity of matching temporal frames while repelling non-matching frames: 
\begin{equation}
\mathcal{L}_{CL} = - \frac{1}{N} \sum_{i=1}^{N} \log \frac{\exp \left( \langle \hat{z}_{ca}^{(i)}, f_{av}^{(i)} \rangle / \tau \right)}{\sum_{j=1}^{N} \exp \left( \langle \hat{z}_{ca}^{(i)}, f_{av}^{(j)} \rangle / \tau \right)},
\end{equation} 
where $\hat{z}_{ca}^{(i)}$ and $f_{av}^{(i)}$ represent the $i$-th frame of the flattened $\hat{\mathcal{Z}}_{ca}$ and $\mathcal{F}_{av}$ respectively, $\langle \cdot, \cdot \rangle$ denotes the dot product, and $\tau$ is the temperature hyperparameter.

\textbf{Semantic Regularization.} 
 To guarantee pronunciation correctness, we apply a Connectionist Temporal Classification (CTC) $\mathcal{L}_{ctc}$ loss  directly to the final hidden features $\mathcal{Z}^l_{out}$ (\ie, the output of Eq.~\eqref{context_MHCA}). 
 It encourages the model to retain more linguistic information in predicted hidden states. 
By jointly optimizing these objectives within the JSAR framework, the model effectively synchronizes the context with the acoustic dynamics and maintains semantic consistency, stably anchoring the flow matching trajectory.

\subsection{Optimal-Transport Conditional
Flow Matching Objective}
\label{sec:fm_Ob}
The objective of the OT-CFM is to minimize the Mean Squared Error (MSE) between the predicted vector field by CoSync-DiT and the target vector field:
\begin{equation}
\mathcal{L}_{fm} = \mathbb{E}_{t, q(x_1), p(x_0)} [ || v_\theta(x_t | t, \mathcal{H}_{m}, \mathcal{X}_{lip}, \mathcal{H}_{text}) - (x_1 - x_0) ||^2 ],
\end{equation}
where $t \sim \mathcal{U}[0, 1]$ is the timestep, and $x_1 \sim q(x_1)$ denotes the target mel-spectrogram latent. 
$x_t$ represents the  intermediate noise latent. 
$v_\theta$ represents the vector field predicted by the proposed CoSync-DiT with model parameters $\theta$, which is progressively conditioned on the  $\mathcal{H}_{m}$, $\mathcal{X}_{lip}$, and $\mathcal{H}_{text}$.

\subsection{Acoustic-Semantic Classifier-Free Guidance}
\label{sec:cfg}
In conditional flow matching, classifier-free guidance (CFG)~\cite{ho2022classifier} is typically employed to amplify the influence of the conditioning signals. 
Driven by acoustic and semantic prior, our CFG focuses on explicitly decoupling these conditions. 
Let $\mathcal{C}$ denote the joint condition comprising both acoustic and semantic, and $\varnothing$ denote the unconditional prior. 
The modified vector field is formulated as: 
\begin{equation}
    v_{t,cfg}= v_t(x_t, \mathcal{C}) + \lambda_{a}
    \cdot(v_{t}(x_t, \mathcal{C})-v_t(x_t, \mathcal{H}_{m})) 
     + \lambda_{s} 
    \cdot(v_{t}(x_t, \mathcal{H}_{m})-v_t(x_t,\varnothing )), 
\label{final_cfg}
\end{equation}
where $\lambda_{a}$ and $\lambda_{s}$ are the acoustic and semantic guidance scales, enabling a highly controllable dubbing generation.

\section{Experiments}

\subsection{Datasets and Experimental Setup}

In this paper, we conduct experiments on three dubbing datasets, encompassing diverse in-the-wild scenarios (\eg, live action movies, vlogs, and dramas) as well as traditional scenarios. 
Tab.~\ref{tab_1_Setting_CleanExplanation} summarizes the evaluation settings:
(1) \textbf{Setting 1}: ground-truth speech is used as reference;
(2) \textbf{Setting 2}: a different clip from the same speaker serves as reference;
(3) \textbf{Zero-shot}: both unseen voice and unseen video are evaluated by out-of-domain dubbing dataset.

\smallskip \noindent\textbf{Chemistry Lecture (Chem).}
This single-speaker dataset~\cite{prajwal2020learning} features a chemistry teacher delivering educational classroom lectures. It comprises short video segments sourced from YouTube and encompasses approximately nine hours of total video content. Following the standard sentence-level pre-processing~\cite{hu2021neural}, the dataset yields 6,132 training samples and 196 testing samples for  dubbing.

\smallskip \noindent\textbf{CelebV-Dub.}
Built upon the foundations of CelebV-HQ~\cite{HaoZhuCelebV} and CelebV-Text~\cite{JianhuiCelebV}, this dataset aggregates content from diverse sources such as vlogs and dramas. It serves as a distinctly challenging benchmark by featuring unconstrained real-world settings alongside rich emotional variations. Following the official division~\cite{sung2025voicecraft}, this dataset provides 79,933 training samples and 213 testing samples.

\smallskip \noindent\textbf{CinePile-Dub.}
Derived from the original CinePile~\cite{rawal2024cinepile} dataset, CinePile-Dub is specifically curated from professional live-action movie productions. It features high emotional intensity and expressive acting-specific prosody characteristic of authentic cinematic environments. Specifically, it comprises 160 professional movie video clips exclusively reserved to rigorously evaluate dubbing performance in a zero-shot setting.

\begin{table}[!t]
  \centering \caption{Overview of evaluation configurations across three experimental settings in various dubbing benchmarks.  
  } 
  \resizebox{0.8\linewidth}{!}
  {
    \begin{tabular}{llccccccc}
    \toprule
    Dataset & Speaker & Main Scenes & Experiment Setting  \\
    \midrule
    CelebV-Dub~\cite{sung2025voicecraft}& Multi-Speakers &  Vlogs, Dramas & Setting 1 \& 2 \\
    Chem~\cite{hu2021neural} & Single-Speaker & Teaching Video  & Setting 1 \& 2  \\
    \midrule
    CinePile-Dub~\cite{rawal2024cinepile} & Multi-Speakers &  Live Action Movies  & Zero-shot Setting \\
    \bottomrule
    \end{tabular}
    }   \label{tab_1_Setting_CleanExplanation}
\end{table}

\subsection{Evaluation Metrics}
We adopt several authority metrics to comprehensively evaluate various aspects of generated dubbing.

\smallskip \noindent\textbf{WER} measures pronunciation clarity by calculating the word error rate derived from an automatic speech recognition model, whisper-large-V3 ~\cite{whisper}.

\smallskip \noindent\textbf{EMOSIM} measures emotion similarity. 
We compute the cosine similarity between the synthesized speech and the ground truth by the Emotion2Vec model~\cite{Ziyangemotion2vec}, which is a universal speech emotion representation model. 

\smallskip \noindent\textbf{SPKSIM} measures speaker similarity. 
In contrast to the prior metric in~\cite{LiangPAMI}, \ie, speaker encoder cosine similarity (SECS) by GE2E model~\cite{WanGeneralized}, we adopt the WavLM-TDNN model~\cite{SanyuanWavLM}, which is widely adopted in speaker verification to ensure a reliable measure~\cite{sung2025voicecraft, GaoxiangFlowDubber}.

\smallskip \noindent\textbf{Sync-KL}
also known as duration divergence~\cite{zhang2025instructdubber}.
Since visual-based alignment metrics (\eg, AVSync~\cite{AlignDiT} and LSE~\cite{ChungOutofTime}) still suffer from the complex visual changes and perturbations in in-the-wild scenarios, Sync-KL offers a more reliable manner by directly quantifying the duration discrepancy between GT and generated speech, leveraging the fact that GT speech is naturally synchronized with video.
We will provide further comparisons of AV-Sync in the supplementary materials.

\smallskip \noindent\textbf{DNSMOS} is used to objectively evaluate speech quality by approximating subjective human ratings (Mean Opinion Score, MOS). 
DNSMOS~\cite{reddy2021dnsmos} (Deep Noise Suppression MOS) assesses speech clarity and background noise cleanliness. 

\smallskip \noindent\textbf{MOS-N and MOS-S}. 
We will provide human subjective rating results in supplementary materials, including MOS-N (Naturalness) and MOS-S (Similarity).

\subsection{Implementation Details}

The input and output dimensions of the unified projection layer are 712 and 1,024, respectively. 
ConvPosition is used to inject positional information into the unified projection layer with a kernel size of 31 and 16 groups. CoSync-DiT consists of 22 layers. The multi-head self-attention and multi-head cross-attention layers both have a hidden dimension of 1,024 with 16 attention heads. 
Lip regions are resized to $96 \times 96$ and processed by a pre-trained AV-HuBERT~\cite{AVHUBERT} model to extract 1,024-dimensional embeddings. 
The text encoder comprises a stack of four ConvNeXt V2 blocks with a hidden dimension of 512. 
In JSAR, the temperature parameter $\tau$ is set to 0.07. 
The CTC projection layer within JSAR includes two temporal downsampling layers with Mish activations to map the 1,024-dimensional features into a 2,547-dimensional space. 
The final projection layer maps the 1,024-dimensional features to a predicted 100-dimensional vector field. 
During inference, the predicted vector field is used to solve the ODE from Gaussian noise to the target mel-spectrogram using an Euler solver with 32 function evaluations. 
We optimize the model using the AdamW optimizer~\cite{loshchilov2017decoupled} with momentum parameters $\beta_1=0.9$ and $\beta_2=0.999$. 
The decoupled weight decay is set to 0.01, and the epsilon term is $1 \times 10^{-8}$. 
A random 70\% to 100\% span of mel-spectrogram frames is masked with span length $\eta$.

\subsection{Performance Comparison with SOTA Methods}

\begin{table*}[!t]
  \centering
    \caption{
    Compared with SOTA methods on the Chem benchmark under Setting 1. 
    }
  \resizebox{1.0\linewidth}{!}
  {
    \begin{tabular}{c|ccccc}
    \hline
    Methods
    & SPKSIM(\%)$\uparrow$ 
    & WER(\%)$\downarrow$ 
    & EMOSIM(\%)$\uparrow$ 
    & Sync-KL$\downarrow$
    & DNSMOS$\uparrow$
    \\ 
    \midrule
    GT  & 100.00  & 3.85 & 100.00 & 0.00 & 3.86 \\
    \midrule
    HPMDubbing~\cite{cong2023learning} (CVPR'23) & 57.05 & 17.52 & 78.80 & 0.477 & 3.34 \\
    StyleDubber~\cite{cong2024styledubber} (ACL'24) &  61.87 & 10.62 & 80.95 & 0.440 & 3.39  \\
    EmoDubber~\cite{cong2024emodubber} (CVPR'25) &  75.60  & 11.86  & 85.10   & 0.420  & 3.70 \\
    FlowDubber~\cite{GaoxiangFlowDubber} (MM'25) &  75.37  & 11.51  & 83.47   & 0.417  & 3.66  \\
    HD-Dubber~\cite{LiangPAMI}(TPAMI'25) & 64.78  & 14.23 & 79.59 &  0.410 & 3.60  \\
    AlignDiT~\cite{AlignDiT} (MM'25)& 72.73  & 12.39 &  86.28  &  0.349 & 3.80 \\
    Produbber~\cite{zhang2025produbber} (CVPR'25) &  38.55 & 9.45  & 77.93 & 0.464  &3.62 \\
    InstructDub~\cite{zhang2025instructdubber} (AAAI'26) & 42.10 &   8.86  & 80.99  &  0.431 & 3.82 \\
    \midrule
     Ours & \textbf{81.84} & \textbf{7.04}  &  \textbf{87.84}    & \textbf{0.289}   & \textbf{3.83}  \\
    \bottomrule
    \end{tabular}}
  \label{result_Chem_s1}%
\end{table*}%

\begin{table*}[!t]
  \centering
    \caption{
    Compared with SOTA methods on the Chem benchmark under Setting 2. 
    }
  \resizebox{1.0\linewidth}{!}
  {
    \begin{tabular}{c|ccccc}
    \hline
    Methods
    & SPKSIM(\%)$\uparrow$ 
    & WER(\%)$\downarrow$ 
    & EMOSIM(\%)$\uparrow$ 
    & Sync-KL$\downarrow$
    & DNSMOS$\uparrow$
    \\ 
    \midrule
    GT  & 73.01  & 3.85 & 100.00 & 0.00 & 3.86 \\
    \midrule
    HPMDubbing~\cite{cong2023learning} (CVPR'23) & 44.67 & 18.48 & 77.49 & 0.458  & 3.33 \\
    StyleDubber~\cite{cong2024styledubber} (ACL'24) &  50.00  & 11.98  & 79.09  & 0.426  & 3.34  \\
    EmoDubber~\cite{cong2024emodubber} (CVPR'25) &  67.53  & 12.01  & 79.41 & 0.427  & 3.71 \\
    FlowDubber~\cite{GaoxiangFlowDubber} (MM'25) &  60.61 & 15.24 & 78.95 & 0.411   & 3.66  \\
    HD-Dubber~\cite{LiangPAMI}(TPAMI'25) & 52.70 & 14.80 & 77.77 & 0.430 & 3.59  \\
    AlignDiT~\cite{AlignDiT} (MM'25)& 64.08 & 13.28 & 83.76 &  0.349 & 3.82 \\
    Produbber~\cite{zhang2025produbber} (CVPR'25) &  26.87  & 11.69 & 77.42 & 0.412  &3.53 \\
    InstructDub~\cite{zhang2025instructdubber} (AAAI'26) & 35.31 & 8.46  &  78.48  & 0.433& 3.83 \\
    \midrule
     Ours & \textbf{72.29} & \textbf{8.43}  &  \textbf{87.06}    & \textbf{0.288}   & \textbf{3.84}  \\
    \bottomrule
    \end{tabular}}
  \label{result_Chem_s2}%
\end{table*}%

\smallskip \noindent\textbf{Results on the Chem (Setting 1)}. 
As shown in Tab.~\ref{result_Chem_s1}, our proposed CoSync-DiT achieves the best performance across all evaluated metrics. 
Under Setting 1, it attains 81.84\% in SPKSIM, surpassing the second-best EmoDubber by an absolute margin of 6.24\%, demonstrating exceptional capability in high-fidelity speaker preservation. 
Furthermore, our method achieves 87.84\% in EMOSIM and the lowest WER of 7.04\%, validating the effectiveness of our time-aware context aligning and semantic consistency constraints for precise pronunciation modeling. 
Finally, the lowest Sync-KL score of 0.289 and the highest DNSMOS of 3.83 among all dubbing baselines further confirm its superior fine-grained audio-visual synchronization and premium acoustic clarity.

\smallskip \noindent\textbf{Results on the Chem (Setting 2)}. 
As presented in Tab.~\ref{result_Chem_s2}, we further evaluate the models under the distinctly more challenging Setting 2 scenario. Despite the increased difficulty, our proposed CoSync-DiT consistently maintains its superiority across all evaluated metrics. Specifically, it achieves 72.29\% in SPKSIM, outperforming the second-best EmoDubber by an absolute margin of 4.76\% and remarkably approaching the Ground Truth upper bound of 73.01\%. 
This demonstrates the exceptional robustness of our acoustic style adapting phase in capturing complex unseen timbres. 
Furthermore, our method yields the highest EMOSIM of 87.06\% and the lowest WER of 8.43\%. 
Finally, CoSync-DiT attains the exceptionally low Sync-KL score of 0.288 and a remarkable DNSMOS of 3.84, confirming that our fine-grained visual calibrating module effectively guarantees strict audio-visual synchronization without sacrificing premium acoustic fidelity.

\smallskip \noindent\textbf{Results on the CelebV-Dub (Setting 1)}. Results are reported in Tab.~\ref{result_CelebVDub_S1}. Compared with relatively controlled datasets like Chem, CelebV-Dub contains diverse in-the-wild content such as vlogs and dramas with rich emotional variations, posing a distinctly more challenging evaluation. Despite these complexities, our proposed CoSync-DiT successfully achieves state-of-the-art performance across all evaluated metrics. 
Specifically, it attains 65.21\% in SPKSIM, outperforming the strongest baseline AlignDiT by an absolute margin of 5.50\%, which confirms the reliability of our acoustic style adapting phase in handling highly expressive speech. 
Furthermore, our method obtains the highest EMOSIM of 84.61\% alongside an exceptionally low WER of 4.29\%, closely approaching the Ground Truth limit of 4.15\%. 
Finally, securing the lowest Sync-KL score of 0.392 and the highest DNSMOS of 3.46 proves that our framework consistently delivers precise audio-visual synchronization and premium acoustic fidelity even in challenging dubbing scenarios.

\begin{table}[!t]
  \centering
    \caption{
    Compared with SOTA Dubbing methods on the CelebV-Dub dataset under Setting 1. 
    Unlike single dubbing scenario (\eg, Chem), 
    CelebV-Dub includes diverse video content (vlogs and dramas) with highly expressive speech. 
    }
  \resizebox{1.0\linewidth}{!}
  {
    \begin{tabular}{c|ccccc}
    \hline
    Methods
    & SPKSIM(\%)$\uparrow$ 
    & WER(\%)$\downarrow$ 
    & EMOSIM(\%)$\uparrow$ 
    & Sync-KL$\downarrow$ 
    & DNSMOS$\uparrow$ 
    \\ 
    \midrule
    GT  & 100.00 & 4.15  &  100.00  & 0.000 & 3.38 \\
    \midrule
    HPMDubbing~\cite{cong2023learning} (CVPR'23) & 17.26 & 21.99 & 73.01   & 0.441 & 2.47 \\
    StyleDubber~\cite{cong2024styledubber}(ACL'24) & 26.39 & 10.79  & 79.38  &0.415  & 2.65  \\
    EmoDubber~\cite{cong2024emodubber} (CVPR'25) & 22.08  & 13.44  & 76.59   & 0.407 & 3.26 \\
    FlowDubber~\cite{GaoxiangFlowDubber} (MM'25) & 22.24 & 13.95 & 76.76  & 0.423 & 3.27 \\
    Produbber~\cite{zhang2025produbber} (CVPR'25) & 26.97 & 5.18  & 73.49  & 0.421 & 3.12 \\
    HD-Dubber~\cite{LiangPAMI} (TPAMI'25) & 10.96 & 66.30 & 19.73  & 0.410 & 3.01 \\
    AlignDiT~\cite{AlignDiT} (MM'25) & 59.71 &  9.48  &  84.54 & 0.402 & 3.45\\
    InstructDub~\cite{zhang2025instructdubber} (AAAI'26) & 29.12  & 6.13   & 75.57  & 0.429   & 3.16 \\
    \midrule
     Ours & \textbf{65.21}  &  \textbf{4.29}    &  \textbf{84.61}    &  \textbf{0.392}  &  \textbf{3.46} \\
    \bottomrule
    \end{tabular}
}
  \label{result_CelebVDub_S1}
\end{table}%

\begin{table}[!t]
  \centering
    \caption{
    Compared with SOTA methods on the CelebV-Dub dataset under Setting 2.  
    }
  \resizebox{1.0\linewidth}{!}
  {
    \begin{tabular}{c|ccccc}
    \hline
    Methods
    & SPKSIM(\%)$\uparrow$ 
    & WER(\%)$\downarrow$ 
    & EMOSIM(\%)$\uparrow$ 
    & Sync-KL$\downarrow$ 
    & DNSMOS$\uparrow$ 
    \\ 
    \midrule
    GT  & 66.13 & 4.15  &  100.00 & 0.000 & 3.38  \\
    \midrule
    HPMDubbing~\cite{cong2023learning} (CVPR'23) & 12.52 & 23.64  & 69.25  & 0.447    &  2.47 \\
    StyleDubber~\cite{cong2024styledubber} { (ACL'24)} & 19.42 & 7.03 & 74.87 & 0.405 & 2.63  \\
    EmoDubber~\cite{cong2024emodubber} (CVPR'25) & 18.78 & 16.37  & 76.30 & 0.422 & 3.30 \\
    FlowDubber~\cite{GaoxiangFlowDubber} (MM'25) &  19.32 & 14.94 & 74.18 & 0.420 & 3.31 \\
    Produbber~\cite{zhang2025produbber} (CVPR'25)  & 23.13 & 6.41 & 71.38 & 0.432 & 3.14  \\
    HD-Dubber~\cite{LiangPAMI}(TPAMI'25) & 9.69  & 64.71  & 16.86 & 0.400 & 3.00 \\
    AlignDiT~\cite{AlignDiT} { (MM'25)} & 49.49 & 13.18 &  79.69 & 0.413 & \textbf{3.47} \\
    InstructDub~\cite{zhang2025instructdubber} (AAAI'26) & 22.85  & \textbf{5.64}   & 74.03 & 0.434 & 3.18  \\
    \midrule
     Ours & \textbf{53.44}  &  {6.39}    &  \textbf{80.29}    &  \textbf{0.381}  &  \textbf{3.47} \\
    \bottomrule
    \end{tabular}
}
  \label{result_CelebVDub_S2}
\end{table}%

\smallskip \noindent\textbf{Results on the CelebV-Dub (Setting 2)}. 
Finally, we evaluate the models on the CelebV-Dub dataset under Setting 2. 
As shown in Tab.~\ref{result_CelebVDub_S2}, our CoSync-DiT achieves the highest SPKSIM of 53.44\% and EMOSIM of 80.29\%, demonstrating robust acoustic style adapting capabilities for complex unseen voices. While InstructDub attains a marginally lower Word Error Rate of 5.64\%, it severely compromises speaker identity preservation by yielding a remarkably low SPKSIM of 22.85\%. In contrast, our method maintains a highly competitive WER of 6.39\% while simultaneously establishing the best audio-visual synchronization with a Sync-KL of 0.381. Furthermore, achieving the highest DNSMOS score of 3.47 confirms that our dubbing method avoids the pitfall of single-metric over-optimization, providing the most comprehensive and balanced  dubbing quality.

\begin{table}[!t]
  \centering
  \caption{Zero-shot main experiment results on CinePile-Dub dataset (Out-of-Domain). All models are trained on the training set of the basic dubbing dataset CelebV-Dub.} 
  \resizebox{1.0\linewidth}{!}
  {
    \begin{tabular}{c|ccccc}
    \toprule
    Methods
    & SPKSIM(\%)$\uparrow$ 
    & WER(\%)$\downarrow$ 
    & EMOSIM(\%)$\uparrow$
    & Sync-KL$\downarrow$
    & DNSMOS$\uparrow$
    \\ 
    \midrule
     GT & 100.00 & 2.56 & 100.00  & 0.00 &  3.15 \\ 
     \midrule
     HPMDubbing~\cite{cong2023learning}(CVPR’23) & 42.72 & 99.20  & 61.22  & 0.391 & 3.23 \\
    StyleDubber~\cite{cong2024styledubber}(ACL’24) & 41.65 & 74.96  &  63.88  &0.362  & 2.89 \\
    HD-Dubber~\cite{LiangPAMI}](TPAMI’25)  &  5.84 & 42.75  & 56.32   & 0.337  & 2.58\\
    EmoDubber~\cite{cong2024emodubber}(CVPR’25) &  23.10 & 49.57  & 70.08 &  0.337 & 3.05  \\
    FlowDubber~\cite{GaoxiangFlowDubber}(MM’25)  &  22.68 & 54.41  &  67.39 & 0.354 &  3.04 \\
    AlignDiT~\cite{AlignDiT}(MM’25)  & 58.90  &20.98& 77.39&  0.342 & 3.36\\
    ProDubber~\cite{zhang2025produbber}(CVPR’25) & 28.86 & 5.25  & 70.35  & 0.335 &  2.93\\
    InstructDub~\cite{zhang2025instructdubber}(AAAI’26) & 27.61 &   \textbf{4.61} & 65.97  &0.370  & 2.95 \\
    \midrule
     Ours  & \textbf{60.04} &  {5.59}    &  \textbf{77.41}  &    \textbf{0.332}  &  \textbf{3.40}\\ 
    \bottomrule
    \end{tabular}%
    } 
  \label{Zero-shot-CinePile}%
\end{table}

\smallskip \noindent\textbf{Results on the CinePile-Dub (Zero-shot Setting).} 
To assess true zero-shot generalization, we conduct an out-of-domain evaluation where models trained exclusively on CelebV-Dub are tested on CinePile-Dub, a highly challenging real-world live-action movie dataset. 
As presented in Tab.~\ref{Zero-shot-CinePile}, our proposed CoSync-DiT achieves the highest SPKSIM of 60.04\% and EMOSIM of 77.41\%. 
This confirms the powerful cross-domain extrapolation capacity of the proposed method when confronted with entirely novel cinematic environments. 
Although InstructDub attains the lowest Word Error Rate of 4.61\%, it experiences a catastrophic collapse in speaker identity preservation by yielding a poor SPKSIM of 27.61\%. 
Conversely, our method maintains a highly competitive WER of 5.59\% while simultaneously establishing the best speaker similarity 60.04\% and audio-visual synchronization with a leading Sync-KL score of 0.332. 

\begin{figure}[!tb]
  \centering
  \includegraphics[width=1.0\linewidth]{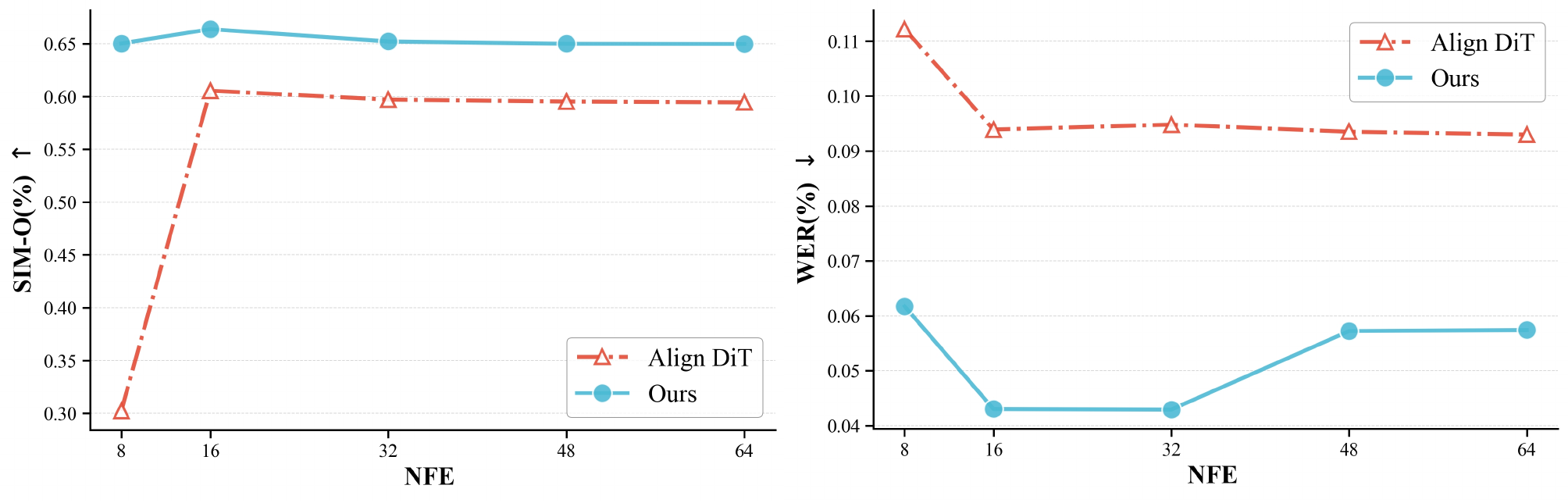}
  \caption{
  Performance comparison between the proposed method and the SOTA baseline under the different Number of Function Evaluations (NFE) steps.  
  }
  \label{fig:NFE}
\end{figure}

\subsection{Ablation Study}

To validate the effectiveness of our core designs, we conduct a comprehensive ablation study on the CelebV-Dub dataset under Setting 2. 
As reported in Tab.~\ref{Ab_Study}, removing the acoustic style adapting phase triggers a severe collapse in speaker similarity, plummeting the SPKSIM from 53.44\% to 19.64\%. 
This confirms its indispensable role in capturing and preserving the target timbre. 
When the fine-grained visual calibrating module is omitted, the Sync-KL metric clearly degrades to 0.419, indicating its absolute necessity for establishing strict lip-sync. 
Furthermore, we find that the absence of the time-aware context aligning phase exerts the most profound impact on overall synchronization, as its core function relies on re-retrieving textual content conditioned on previously integrated information to adjust articulation. 
Finally, we dissect the JSAR mechanism. 
Eliminating the semantic consistency constraint primarily damages the pronunciation accuracy, dropping the temporal consistency constraint strictly deteriorates the audio-visual synchronization, and discarding the entire JSAR framework compounds both errors.

\begin{table}[!t]
  \centering
  \caption{Results of ablation study on CelebV-Dub dataset under Setting 2.} 
  \resizebox{1.0\linewidth}{!}
  {
    \begin{tabular}{c|ccccc}
    \toprule
    Methods
    & SPKSIM (\%) $\uparrow$ 
    & WER (\%) $\downarrow$ 
    & EMOSIM (\%) $\uparrow$
    & Sync-KL $\downarrow$
    & DNSMOS $\uparrow$
    \\ 
     \midrule
     w/o Style Adapting & 19.64 & 6.84
 & 77.24
 & 0.385
  & 3.38 \\
    w/o Visual Calibrating & 53.25 & 6.40 & 80.17 & 0.419 & 3.45 \\
   w/o Context Aligning & 52.75  & 7.39 & 80.04 & 0.446 & 3.44 \\
    w/o JSAR & 51.30 & 8.72 & 80.14 & 0.431 & 3.39 \\
    w/o Semantic Consistency & 52.34 & 8.39 &  80.19& 0.392 & 3.42  \\
    w/o Temporal Consistency & 51.37 & 6.58 &  80.24&  0.425& 3.44  \\
    \midrule
     Full model  & \textbf{53.44}  &  \textbf{6.39}    &  \textbf{80.29}    &  \textbf{0.381}  &  \textbf{3.47}\\ 
    \bottomrule
    \end{tabular}%
    } 
  \label{Ab_Study}%
\end{table}

\begin{figure}[!tb]
  \centering
  \includegraphics[width=0.9\linewidth]{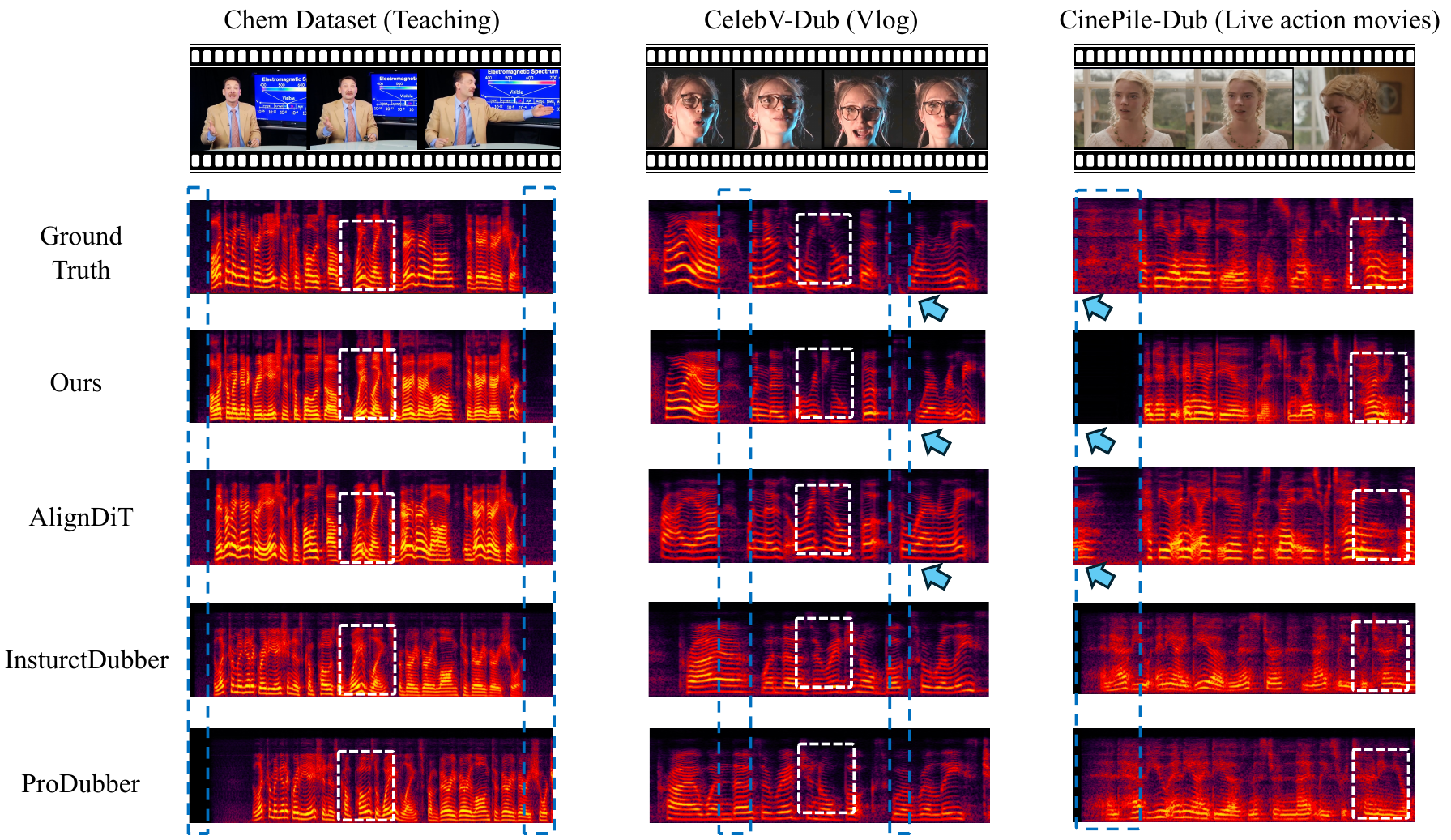}
  \caption{
  Visual comparison of the generated mel-spectrograms. The blue arrows highlight specific regions requiring attention for audio-visual synchronization.  
  }
  \label{fig:Visual}
\end{figure}

\subsection{Comprehensive Analysis of Generative Robustness} 
To comprehensively evaluate the inference efficiency and generation stability of our approach against the state-of-the-art baseline, we conduct a comparative analysis under various Number of Function Evaluations (NFE) settings. As illustrated in Fig.~\ref{fig:NFE}, AlignDiT suffers a catastrophic performance collapse at extremely low sampling steps, evidenced by its SIM-O metric plummeting to approximately 0.30 at an NFE of 8. In stark contrast, our method exhibits remarkable robustness by maintaining a consistently high SIM-O of over 0.65 across all evaluated NFE configurations. Furthermore, CoSync-DiT continuously secures a significantly lower Word Error Rate than AlignDiT at every corresponding step, achieving its optimal pronunciation clarity around an NFE of 16 to 32. 
These results compellingly demonstrate that our method establishes a robust dubbing system, unlocking highly efficient few-step inference without sacrificing acoustic or synchronization quality.

\subsection{Qualitative Comparison}
We visualize the mel-spectrograms of the ground truth and the generated speech from different models in Fig.~\ref{fig:Visual} (additional results are provided in the Appendix). 
The blue dashed regions and arrows highlight critical temporal intervals for audio-visual synchronization, while the white dashed boxes emphasize fine-grained pronunciation differences. While most baseline methods perform adequately on the controlled Chem dataset, they degrade sharply on more challenging benchmarks. Specifically in vlog and cinematic scenarios, models like InstructDubber and ProDubber completely fail to maintain proper audio-visual sync. Furthermore, AlignDiT exhibits noticeable temporal shifts and little distortions within the arrow-indicated regions. Conversely, our proposed CoSync-DiT synthesizes highly robust mel-spectrograms that most closely match the ground truth across both boundaries and details. 

\section{Conclusion} 
In this paper, we propose CoSync-DiT, a flow matching-based framework for movie dubbing, which sequentially executes acoustic style adapting, fine-grained visual calibrating, and time-aware context aligning to explicitly guide the noise-to-speech generative trajectory. 
Furthermore, we introduce the Joint Semantic and Alignment Regularization (JSAR) mechanism to simultaneously enforce frame-level temporal consistency on the contextual outputs and semantic consistency on the flow hidden states, guaranteeing reliable pronunciation during the alignment phase. 
Extensive experiments on standard benchmarks and in-the-wild datasets demonstrate that our method achieves state-of-the-art performance. 

\clearpage
\renewcommand{\thesection}{\Alph{section}}
\setcounter{section}{0} %

\section*{Appendix}
This appendix provides the following extra content:
\begin{itemize}
\item Appendix~\ref{sec:AVSync} provides additional quantitative results using the AVSync metric.
\item Appendix~\ref{sec:zero-shot} presents experimental results under an extreme configuration combining Zero-shot and Setting 2.
\item Appendix~\ref{sec:lrs3_aligndit} includes comparisons with the official checkpoint of AlignDiT trained on the large-scale LRS3 dataset.
\item Appendix~\ref{sec:morevisual} provides further qualitative visualizations and mel-spectrogram analysis. 
\end{itemize}

We will open-source all detailed experimental settings, source code, and pre-trained weights. 

\section{Synchronization Analysis (AVSync)}
\label{sec:AVSync}

In this section, we expand upon our primary experimental results by introducing the AVSync metric to comprehensively evaluate the effectiveness of our proposed method. 
Similar to Sync-KL~\cite{zhang2025instructdubber}, AVSync is a recently proposed metric specifically designed to assess fine-grained lip synchronization~\cite{AlignDiT}. 
As highlighted in recent work~\cite{yaman2024audio,AlignDiT}, AVSync provides a substantially more accurate evaluation of audio-visual alignment than traditional metrics such as LSE-C and LSE-D~\cite{ChungOutofTime}. 
Mechanistically, AVSync evaluates this synchronization by computing the cosine similarity between the AV-HuBERT~\cite{AVHUBERT} features extracted from the video paired with ground-truth speech and those extracted from the video paired with the synthesized speech.

\begin{table}[h]
  \centering
    \caption{
    Compared with SOTA Dubbing methods on the CelebV-Dub dataset under Setting 1. 
    Unlike single dubbing scenario (\eg, Chem), 
    CelebV-Dub includes diverse video content (vlogs and dramas) with highly expressive speech. 
    }
  \resizebox{1.0\linewidth}{!}
  {
    \begin{tabular}{c|cccccc}
    \hline
    Methods
    & SPKSIM(\%)$\uparrow$ 
    & WER(\%)$\downarrow$ 
    & EMOSIM(\%)$\uparrow$ 
    & Sync-KL$\downarrow$ 
    & DNSMOS$\uparrow$ 
    & AVSync(\%)$\uparrow$ 
    \\ 
    \midrule
    GT  & 100.00 & 4.15  &  100.00  & 0.000 & 3.38 & 100.00 \\
    \midrule
    HPMDubbing~\cite{cong2023learning} (CVPR'23) & 17.26 & 21.99 & 73.01   & 0.441 & 2.47 & 47.49\\
    StyleDubber~\cite{cong2024styledubber}(ACL'24) & 26.39 & 10.79  & 79.38  &0.415  & 2.65  &  21.38 \\
    EmoDubber~\cite{cong2024emodubber} (CVPR'25) & 22.08  & 13.44  & 76.59   & 0.407 & 3.26 &32.94 \\
    FlowDubber~\cite{GaoxiangFlowDubber} (MM'25) & 22.24 & 13.95 & 76.76  & 0.423 & 3.27 & 33.40\\
    Produbber~\cite{zhang2025produbber} (CVPR'25) & 26.97 & 5.18  & 73.49  & 0.421 & 3.12 & 15.96 \\
    HD-Dubber~\cite{LiangPAMI} (TPAMI'25) & 10.96 & 66.30 & 19.73  & 0.410 & 3.01 & 18.83 \\
    AlignDiT~\cite{AlignDiT} (MM'25) & 59.71 &  9.48  &  84.54 & 0.402 & 3.45 & 49.05 \\
    InstructDub~\cite{zhang2025instructdubber} (AAAI'26) & 29.12  & 6.13   & 75.57  & 0.429   & 3.16 & 16.74\\
    \midrule
     Ours & \textbf{65.21}  &  \textbf{4.29}    &  \textbf{84.61}    &  \textbf{0.392}  &  \textbf{3.46} & \textbf{65.94} \\
    \bottomrule
    \end{tabular}
    }
  \label{result_CelebVDub_S1_AV}
\end{table}%

\begin{table}[h]
  \centering
    \caption{
    Compared with SOTA methods on the CelebV-Dub dataset under Setting 2.  
    }
  \resizebox{1.0\linewidth}{!}
  {
    \begin{tabular}{c|cccccc}
    \hline
    Methods
    & SPKSIM(\%)$\uparrow$ 
    & WER(\%)$\downarrow$ 
    & EMOSIM(\%)$\uparrow$ 
    & Sync-KL$\downarrow$ 
    & DNSMOS$\uparrow$ 
    & AVSync(\%)$\uparrow$
    \\ 
    \midrule
    GT  & 66.13 & 4.15  &  100.00 & 0.000 & 3.38  & 100.00 \\
    \midrule
    HPMDubbing~\cite{cong2023learning} (CVPR'23) & 12.52 & 23.64  & 69.25  & 0.447    &  2.47 &42.67 \\
    StyleDubber~\cite{cong2024styledubber} { (ACL'24)} & 19.42 & 7.03 & 74.87 & 0.405 & 2.63 & 20.75\\
    EmoDubber~\cite{cong2024emodubber} (CVPR'25) & 18.78 & 16.37  & 76.30 & 0.422 & 3.30 & 32.14\\
    FlowDubber~\cite{GaoxiangFlowDubber} (MM'25) &  19.32 & 14.94 & 74.18 & 0.420 & 3.31 & 32.37\\
    Produbber~\cite{zhang2025produbber} (CVPR'25)  & 23.13 & 6.41 & 71.38 & 0.432 & 3.14 &14.92 \\
    HD-Dubber~\cite{LiangPAMI}(TPAMI'25) & 9.69  & 64.71  & 16.86 & 0.400 & 3.00 &18.45 \\
    AlignDiT~\cite{AlignDiT} { (MM'25)} & 49.49 & 13.18 &  79.69 & 0.413 & \textbf{3.47} & 48.72 \\
    InstructDub~\cite{zhang2025instructdubber} (AAAI'26) & 22.85  & \textbf{5.64}   & 74.03 & 0.434 & 3.18 & 16.16\\
    \midrule
     Ours & \textbf{53.44}  &  {6.39}    &  \textbf{80.29}    &  \textbf{0.381}  &  \textbf{3.47} & \textbf{52.12} \\
    \bottomrule
    \end{tabular}
}
  \label{result_CelebVDub_S2_AV}
\end{table}%

As reported in Tab.~\ref{result_CelebVDub_S1_AV}, our method consistently achieves state-of-the-art results across all metrics after incorporating the AVSync evaluation. Specifically, our approach obtains an AVSync score of 65.94\%, which outperforms the strongest baseline AlignDiT by an absolute margin of 16.89\%. 
This confirms the effectiveness of our proposed cognitive synchronous diffusion mechanism. 
By progressively guiding the denoising process, our method introduces acoustic style adapting, fine-grained visual calibrating, and time-aware context aligning at different network stages. 
This structured integration effectively prevents feature entanglement during the early sampling steps and establishes a precise temporal alignment between the generated speech and the target lip movements. 
Furthermore, under the challenging Setting 2 scenario of CelebV-Dub (see Tab.~\ref{result_CelebVDub_S2_AV}), although our method marginally trails the powerful TTS-pretrained baseline InstructDubber in Word Error Rate, it decisively surpasses all comparative methods in both speaker similarity and comprehensive synchronization metrics (both Sync-KL and AVSync). 
Consistent with these findings, our method sustains this comprehensive superiority in the out-of-domain movie evaluation (see  Tab.~\ref{Zero-shot-CinePile-Setting1}), confirming its robust generalization capabilities and optimal alignment performance across unseen domains.

\begin{table}[!h]
  \centering
  \caption{Zero-shot results on CinePile-Dub dataset (Out-of-Domain Movie). All models are trained on the training set of the basic dubbing dataset CelebV-Dub.} 
  \resizebox{1.0\linewidth}{!}
  {
    \begin{tabular}{c|cccccc}
    \toprule
    Methods
    & SPKSIM(\%)$\uparrow$ 
    & WER(\%)$\downarrow$ 
    & EMOSIM(\%)$\uparrow$
    & Sync-KL$\downarrow$
    & DNSMOS$\uparrow$
    & AVSync(\%)$\uparrow$
    \\ 
    \midrule
     GT & 100.00 & 2.56 & 100.00  & 0.00 &  3.15 & 100.00 \\ 
     \midrule
     HPMDubbing~\cite{cong2023learning}(CVPR’23) & 42.72 & 99.20  & 61.22  & 0.391 & 3.23 & 17.00\\
    StyleDubber~\cite{cong2024styledubber}(ACL’24) & 41.65 & 74.96  &  63.88  &0.362  & 2.89 & 22.56 \\
    HD-Dubber~\cite{LiangPAMI}](TPAMI’25)  &  5.84 & 42.75  & 56.32   & 0.337  & 2.58 & 19.33 \\
    EmoDubber~\cite{cong2024emodubber}(CVPR’25) &  23.10 & 49.57  & 70.08 &  0.337 & 3.05 & 13.75 \\
    FlowDubber~\cite{GaoxiangFlowDubber}(MM’25)  &  22.68 & 54.41  &  67.39 & 0.354 &  3.04 & 13.58 \\
    AlignDiT~\cite{AlignDiT}(MM’25)  & 58.90  &20.98& 77.39&  0.342 & 3.36 & 31.77\\
    ProDubber~\cite{zhang2025produbber}(CVPR’25) & 28.86 & 5.25  & 70.35  & 0.335 &  2.93 & 21.79 \\
    InstructDub~\cite{zhang2025instructdubber}(AAAI’26) & 27.61 &   \textbf{4.61} & 65.97  &0.370  & 2.95 & 17.65\\
    \midrule
     Ours  & \textbf{60.04} &  {5.59}    &  \textbf{77.41}  &    \textbf{0.332}  &  \textbf{3.40} & \textbf{45.24} \\ 
    \bottomrule
    \end{tabular}%
    } 
  \label{Zero-shot-CinePile-Setting1}%
\end{table}

\section{Robustness Analysis in Challenging Scenarios (both Zero-Shot and Setting 2)}
\label{sec:zero-shot}

Although existing experimental configurations are sufficiently challenging, they primarily cover CelebV-Dub under Setting 1 and Setting 2 alongside a standard zero-shot evaluation on out-of-domain movies. 
In this section, we introduce a significantly more rigorous setting that integrates Setting 2 into the zero-shot evaluation. 
This setup simultaneously removes both the training domain priors and the interference of target speech, ultimately creating a highly demanding evaluation scenario. 
We systematically summarize all configurations in Tab.~\ref{tab_sm_Setting_CleanExplanation}. 
We hope this robust evaluation design promotes the advancement of the reliable visual dubbing field and communities. 
We plan to open-source all detailed experimental settings, source code, and pre-trained weights.

\begin{table}[!h]
  \centering \caption{Summary of the experiment configurations in the supplementary materials. Note that Setting 1 and Setting 2 define the reference audio selection strategy. 
  Setting 1 uses the target speech as the reference audio, while Setting 2 uses an unaligned utterance from the same speaker. 
  The Zero-shot condition indicates whether the target domain data is excluded during the training phase. 
  } 
  \resizebox{0.75\linewidth}{!}
  {
    \begin{tabular}{l|c|c}
    \toprule
    Table  & Reference audio Setting & Zero-shot Setting (OOD Movie)  \\
    \midrule
    Tab.~\ref{result_CelebVDub_S1_AV} &  Setting 1 & $\times$\\
    Tab.~\ref{result_CelebVDub_S2_AV}  & Setting 2  &  $\times$ \\
    \midrule
    Tab.~\ref{Zero-shot-CinePile-Setting1} &  Setting 1 & \checkmark \\
    Tab.~\ref{Zero-shot-CinePile-Setting2} &  Setting 2 & \checkmark \\
    \midrule
    Tab.~\ref{Appendix_S1CelebVDub} &  Setting 1 & \checkmark \\
    Tab.~\ref{Official_result_CelebVDub_S2} &  Setting 2 & \checkmark \\
    \bottomrule
    \end{tabular}
    }   \label{tab_sm_Setting_CleanExplanation}
\end{table}

As reported in Tab.~\ref{Zero-shot-CinePile-Setting2}, the performance of all evaluated methods experiences a noticeable decline under this highly constrained configuration. 
Despite these extreme challenges, our proposed method consistently secures the best performance across all objective metrics. 
Specifically, our approach achieves the highest speaker similarity of 47.24\% and emotional similarity of 70.12\%. 
This demonstrates its superior capability to reliably extract and transfer target vocal characteristics from temporally unaligned reference segments. Furthermore, our model attains the lowest Word Error Rate of 7.53\%, successfully surpassing the strong InstructDubber baseline. 
Regarding synchronization, our method records the most precise Sync-KL of 0.319 and an AVSync score of 31.79\%, outperforming the strongest baseline AlignDiT by a solid absolute margin of 9.29\%. 
These comprehensive results compellingly confirm the exceptional robustness of our CoSync-DiT. 
It proves that our progressive cognitive synchronous architecture can reliably maintain high-fidelity speech generation and strict audio-visual alignment even in the most demanding out-of-domain scenarios.

\begin{table}[!h]
  \centering
  \caption{Zero-shot results on the CinePile-Dub dataset (Out-of-Domain Movie) under Setting 2. 
  All models are trained on the training set of the basic dubbing dataset CelebV-Dub.
  The reference audio is from different segments of the same speaker. } 
  \resizebox{1.0\linewidth}{!}
  {
    \begin{tabular}{c|cccccc}
    \toprule
    Methods
    & SPKSIM(\%)$\uparrow$ 
    & WER(\%)$\downarrow$ 
    & EMOSIM(\%)$\uparrow$
    & Sync-KL$\downarrow$
    & DNSMOS$\uparrow$
    & AVSync(\%)$\uparrow$
    \\ 
    \midrule
     GT & 60.58 & 2.56 & 100.00  & 0.00 &  3.15 & 100.00 \\ 
     \midrule
     HPMDubbing~\cite{cong2023learning}(CVPR’23) & 30.75 & 93.86  &  52.93 & 0.382 & 3.24 &  13.87 \\
    StyleDubber~\cite{cong2024styledubber}(ACL’24) &28.67  &  77.61 & 57.15  & 0.372 & 2.88 & 16.48 \\
    HD-Dubber~\cite{LiangPAMI}](TPAMI’25)  & 7.04 &  64.12 &  47.66 & 0.330 & 2.58 & 18.86 \\
    EmoDubber~\cite{cong2024emodubber}(CVPR’25) & 19.43 & 57.73  & 62.46  & 0.347 &  3.05 & 13.00 \\
    FlowDubber~\cite{GaoxiangFlowDubber}(MM’25) & 19.86 &  55.93 &  63.86 & 0.346 & 3.05 & 13.05 \\
    AlignDiT~\cite{AlignDiT}(MM’25)  & 38.58 & 34.52  & 66.74  & 0.358 &  3.37 & 22.50 \\
    ProDubber~\cite{zhang2025produbber}(CVPR’25) &  23.91 &  8.02 &  62.06 & 0.338 & 2.93 &  20.80 \\
    InstructDub~\cite{zhang2025instructdubber}(AAAI’26) & 23.11 & 7.62  &   61.14 &  0.367 & 2.94 & 16.66 \\
    \midrule
     Ours  & \textbf{47.24} &  \textbf{7.53}    &  \textbf{70.12}  &    \textbf{0.319}  &  \textbf{3.39} & \textbf{31.79} \\ 
    \bottomrule
    \end{tabular}%
    } 
  \label{Zero-shot-CinePile-Setting2}%
\end{table}

\section{Comparison with the Official AlignDiT Checkpoint}
\label{sec:lrs3_aligndit}

To the best of our knowledge, AlignDiT~\cite{AlignDiT} represents one of the most advanced methods in the video-to-speech field. The official AlignDiT model is trained on the massive LRS3 dataset~\cite{LRS3}, which comprises approximately 131,000 utterances and 439 hours of unconstrained video content. 
In our primary experiments, we trained both our proposed method and the AlignDiT baseline on the CelebV-Dub dataset to ensure fairness. 
CelebV-Dub contains 67,549 samples totaling around 85.94 hours. 
Note that CelebV-Dub is currently the most popular and accessible large-scale dubbing dataset. 
Unfortunately, the complete LRS3 dataset is no longer available for public download from its official source. This restriction physically prevents us from training our model directly on the LRS3 dataset. 
Nevertheless, to conduct a comprehensive and fair comparison with the most powerful AlignDiT variant, we introduce an extended evaluation in this section. 
Specifically, we utilize the official pre-trained weights released by the AlignDiT authors to guarantee a rigorous and unbiased assessment. 
We denote this official LRS3-trained model as AlignDiT*. 
To establish a strictly neutral testing ground, we evaluate all models on the out-of-domain CinePile-Dub cinematic dataset. 
This configuration ensures a completely fair zero-shot scenario for all competitors.

\begin{table}[h]
  \centering
    \caption{
     Zero-shot evaluation results on the CinePile-Dub dataset (Out-of-domain Movie) under Setting 1. 
     The asterisk (*) denotes the official AlignDiT checkpoint pre-trained on the large-scale LRS3 dataset. 
     All evaluated models face completely unseen target domains. 
    }
  \resizebox{1.0\linewidth}{!}
  {
    \begin{tabular}{c|cccccc}
    \hline
    Methods
    & SPKSIM(\%)$\uparrow$ 
    & WER(\%)$\downarrow$ 
    & EMOSIM(\%)$\uparrow$ 
    & Sync-KL$\downarrow$ 
    & DNSMOS$\uparrow$ 
    & AVSync(\%)$\uparrow$ 
    \\ 
    \midrule
    GT  & 100.00 & 4.15  &  100.00  & 0.000 & 3.38 & 100.00 \\
    \midrule
    AlignDiT*~\cite{AlignDiT}& 61.51 & 18.35   & 76.33   & 0.338 & 3.25 &  24.03 \\
    AlignDiT~\cite{AlignDiT}& 58.90  &20.98& 77.39&  0.342 & 3.36 & 31.77\\
    \midrule
    Ours  & \textbf{60.04} &  \textbf{5.59}    &  \textbf{77.41}  &    \textbf{0.332}  &  \textbf{3.40} & \textbf{45.24} \\ 
    \bottomrule
    \end{tabular}
    }
  \label{Appendix_S1CelebVDub}
\end{table}%

\begin{table}[h]
  \centering
    \caption{
     Zero-shot evaluation results on the CinePile-Dub dataset (Out-of-domain Movie) under Setting 2. 
     All evaluated models face completely unseen target domains and non-target reference audio. 
     The asterisk (*) denotes the official AlignDiT checkpoint pre-trained on the large-scale LRS3 dataset.  
    }
  \resizebox{1.0\linewidth}{!}
  {
    \begin{tabular}{c|cccccc}
    \hline
    Methods
    & SPKSIM(\%)$\uparrow$ 
    & WER(\%)$\downarrow$ 
    & EMOSIM(\%)$\uparrow$ 
    & Sync-KL$\downarrow$ 
    & DNSMOS$\uparrow$ 
    & AVSync(\%)$\uparrow$ 
    \\ 
    \midrule
    GT  & 100.00 & 4.15  &  100.00  & 0.000 & 3.38 & 100.00 \\
    \midrule
    AlignDiT*~\cite{AlignDiT} & 44.01 &  35.90  & 68.80  & 0.359 & 3.23 &  18.72 \\
    AlignDiT~\cite{AlignDiT} & 38.58 & 34.52  & 66.74  & 0.358 &  3.37 & 22.50  \\
    \midrule
    Ours & \textbf{47.24} &  \textbf{7.53}    &  \textbf{70.12}  &    \textbf{0.319}  &  \textbf{3.39} & \textbf{31.79}  \\
    \bottomrule
    \end{tabular}
    }
  \label{Official_result_CelebVDub_S2}
\end{table}%

\begin{figure}[!t]
  \centering
  \includegraphics[width=0.95\linewidth]{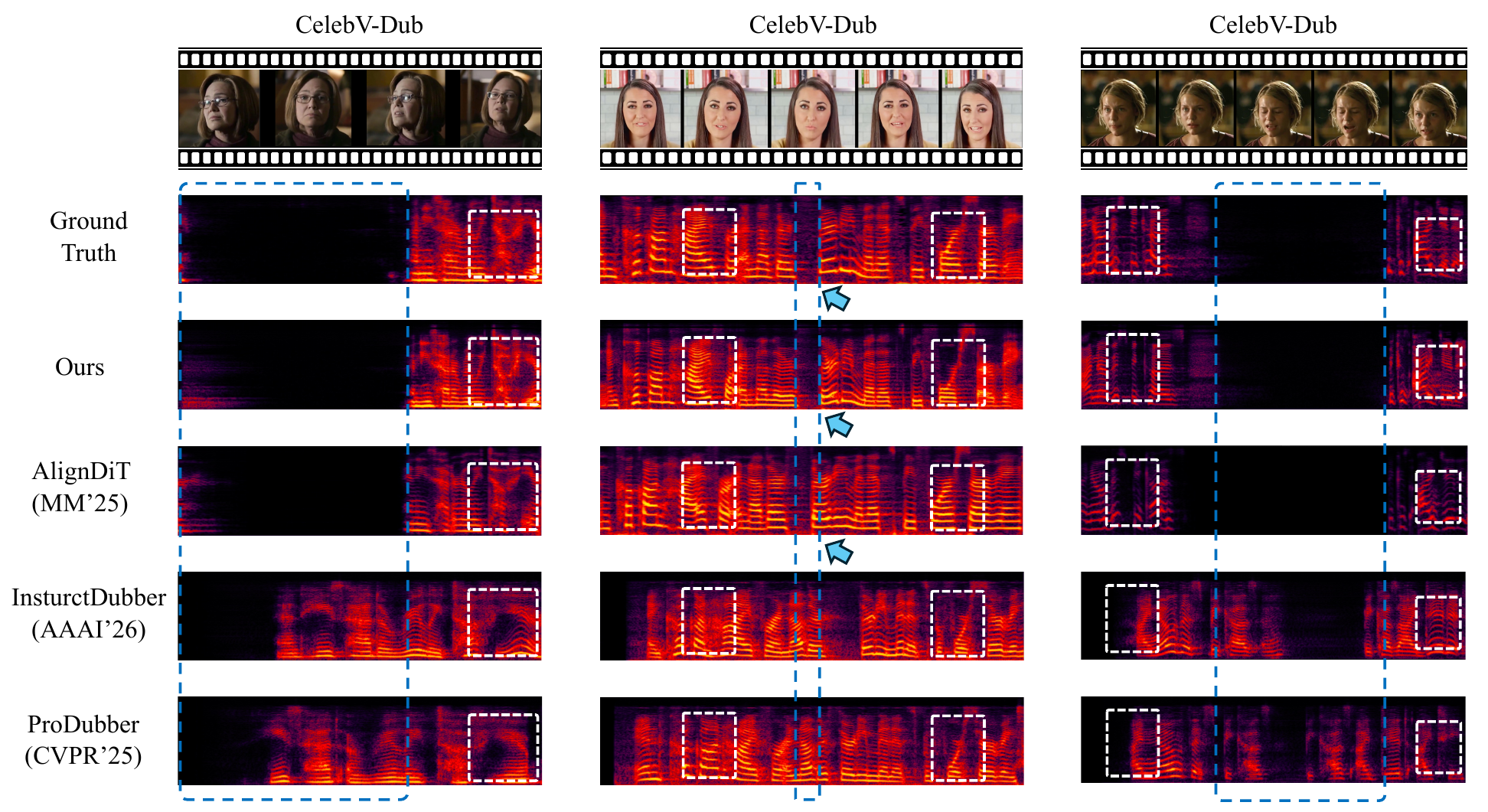}
  \caption{
  Visual comparison of the generated mel-spectrograms with ground truth. 
  The blue and white bounding boxes highlight regions where different models exhibit significant differences in duration and spectrogram details. Furthermore, the blue arrows pinpoint specific temporal regions that require critical attention for evaluating audio-visual synchronization.
  }
  \label{fig:Visual1}
\end{figure}

\begin{figure}[tb]
  \centering
  \includegraphics[width=0.95\linewidth]{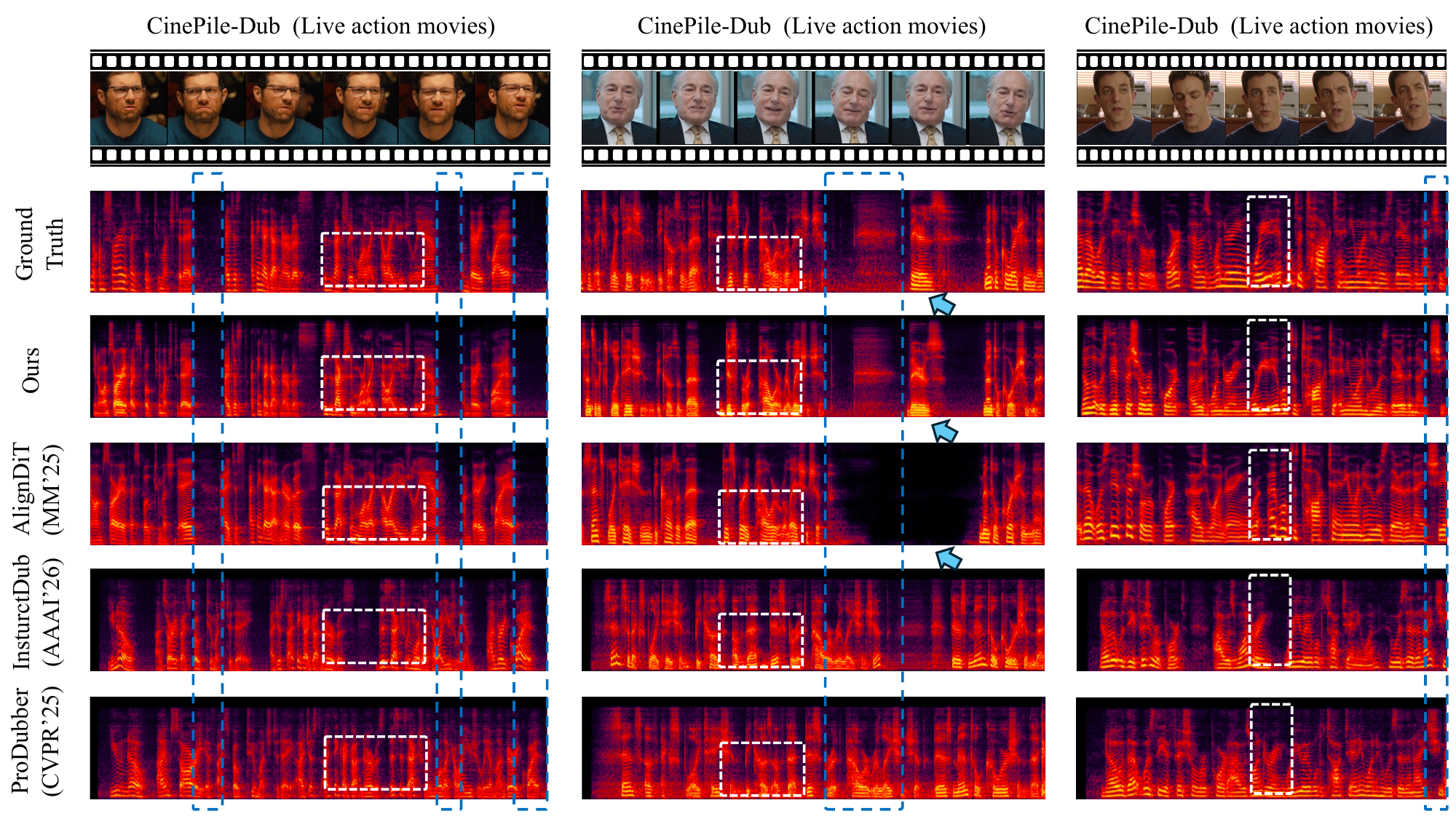}
  \caption{
  Visual comparison of the generated mel-spectrograms with ground truth under zero-shot setting (out-of-domain movie scenario). 
  The blue and white bounding boxes highlight regions where different models exhibit significant differences in duration and spectrogram details. Furthermore, the blue arrows pinpoint specific temporal regions that require critical attention for evaluating audio-visual synchronization. 
  }
  \label{fig:Visual2}
\end{figure}

The comprehensive zero-shot quantitative results are reported in Tab.~\ref{Appendix_S1CelebVDub} (Zero-shot \& Setting1) and Tab.~\ref{Official_result_CelebVDub_S2} (Zero-shot \& Setting2). 
First, we observe that the AlignDiT model trained on CelebV-Dub achieves performance highly comparable to the official AlignDiT* model trained on LRS3. 
This directly confirms that our baseline implementation on CelebV-Dub is solid and does not disadvantage the comparative architecture. 
More importantly, our proposed method consistently outperforms both AlignDiT variants across all objective metrics. 
Even though our model is trained on a dataset approximately one-fifth the size of LRS3, it significantly suppresses the Word Error Rate and achieves much higher AVSync scores in both settings. 
This compellingly demonstrates that our architectural design possesses superior cross-modal alignment capabilities and stronger zero-shot generalization than merely scaling up the training data volume.

\section{More Qualitative Comparison}
\label{sec:morevisual}
In this section, we provide additional visual examples of the generated mel-spectrograms to further evaluate the generated quality. 
As illustrated in Fig.~\ref{fig:Visual1} and Fig.~\ref{fig:Visual2}, our proposed method demonstrates a distinct advantage in preserving both fine-grained acoustic details and audio-visual alignment. 
We specifically highlight the critical temporal regions indicated by the blue arrows. 
For instance, in the middle column of Fig.~\ref{fig:Visual1}, this baseline exhibits a slight temporal shift. 
Similarly, in the middle column of Fig.~\ref{fig:Visual2}, the state-of-the-art AlignDiT baseline completely misses the specific acoustic regions that should be synchronized with the ground truth.  
Conversely, our method, shown in the second row, reconstructs these specific acoustic patterns with strict temporal alignment. 
Furthermore, by observing the white rectangular boxes, our method presents highly clear spectrogram details. This directly reflects the robust timbre preservation and pronunciation clarity of our generated speech. This visual evidence intuitively validates the effectiveness of our method in guaranteeing precise audio-visual synchronization. Additionally, comparing our approach with duration predictor-based methods like InstructDubber and ProDubber reveals a fundamental architectural limitation of those baselines. Although such methods yield high pronunciation clarity, they completely fail to preserve audio-visual alignment in practical dubbing scenarios. As indicated by the blue rectangular boxes, their temporal synchronization is highly inaccurate.

\bibliographystyle{splncs04}
\bibliography{main}

\end{document}